\documentstyle[pre,aps,epsf]{revtex} 

\newcommand{\be}{\begin{equation}}
\newcommand{\ee}{\end{equation}}

\newcommand{\bea}{\begin{eqnarray}}
\newcommand{\eea}{\end{eqnarray}}

\newcommand{\bdm}{\begin{displaymath}}
\newcommand{\edm}{\end{displaymath}}

\begin{document} 
\draft

 \newcommand{\h}{\mbox{\sf h}}

\title{
Domain wall roughening in three dimensional magnets at the depinning transition
}

\author{M. Jost \cite{mjtEmail} and K.~D. Usadel\cite{usadelEmail}}

\address{
Theoretische Tieftemperaturphysik, 
Gerhard-Mercator-Universit\"at Duisburg,\\ 
Lotharstr. 1, 47048 Duisburg, Germany \\
}

\date{09.10.1996}

\maketitle

\begin{abstract}
\newline 
\vspace{-1cm}
\begin{center}
 Accepted for publication in Physica A
\end{center}
The kinetic roughening of a driven interface between three dimensional spin-up and
spin-down domains in a model with non-conserved scalar order parameter and quenched
disorder is studied numerically within a discrete time dynamics at zero temperature.
The exponents characterizing the morphology of the interface are
obtained close to the depinning transition.
\end{abstract}

\section{Introduction} 
A variety of interface roughening models with quenched disorder have been studied 
recently (for an overview see \cite{NATO} and references therein).
These models have in common the occurrence of a so called depinning
transition from a phase where the interface is trapped to a moving phase.
To understand this transition and the morphology of the moving interface
in the presence of quenched noise one may start with the continuum
equation 
\be
 \frac{\partial \h}{\partial t} = \nu \nabla^2 \h +F+ \eta({\bf x},\h({\bf x},t))
\label{ew}
\ee
where $\h({\bf x},t)$ denotes the interface profile, $F$ a homogenous driving force
and $\eta({\bf x},\h({\bf x},t))$ a noise term  which is time independent and depends
only on the local position of the interface. The first term on the right hand side
of this equation which is also known as Edwards-Wilkinson (EW) equation \cite{EW} 
models a surface tension
having a smoothing effect on the interface while the noise term roughens the interface.
As has been shown by Bruinsma and Aeppli \cite{Bruin}, Eq.~(\ref{ew})
arises naturally in magnetic systems where the interface is between a domain with
spins pointing in one direction and another domain with spins pointing in the 
opposite direction. Additionally, it has also been used for describing fluid flow
in porous media \cite{Kop}. \\
Eq.~(\ref{ew}) was modified by Kardar et al. \cite{KPZ}
who added the nonlinear term $\frac{\lambda}{2} (\nabla \h)^2$ with $\lambda$  
proportional to the interface velocity. Such a term and many others arise when
one tries to derive a continuum equation for the position of the interface from a 
Ginzburg-Landau-type free energy functional \cite{KrugSpohn}. 
This approach can be traced back to Ref. \cite{alt} where it was shown that the 
component of the local velocity normal to the domain wall is proportional to its
local curvature. Expressing the position of the wall by a function
$\h({\bf x},t)$
assuming that there are no overhangs it is only a simple mathematical exercise
to derive from this the following equation of motion valid on length
scales large compared to the intrinsic width of the interface,
\be
\frac{1}{\sqrt{g}}\frac{\partial \h}{\partial t} = 
\frac{\nu}{\sqrt{g}^{3/2}} \nabla^2 \h +F+ \eta({\bf x},\h({\bf x},t)),
\ee
where $g=1+(\nabla \h)^2$. The EW and the KPZ equation, respectively, follows by expanding in 
$(\nabla \h)^2$ for nearly flat interfaces.

There exists by now an extensive literature on these nonlinear equations as well
as on various surface or interface models. Numerical studies of these models gave
values for the characteristic critical exponents with rather large scattering. The
question whether these numerical models  fall into one of the two universality
classes defined by the EW equation or the KPZ equation is therefore
not always clear.

In the present paper we add to this ongoing discussion a numerical study on the 
morphology of a domain wall in a three dimensional magnetic medium with quenched
random fields which is driven by an external magnetic field. The magnetic medium
is described by a Ginzburg-Landau-type energy functional and a zero temperature
Langevin dynamics is imposed. We start from this semi-microscopic model
since it is certainly more realistic than an equation of motion for the 
interface which is only obtained after quite a few more or less
plausible approximations.

It has been argued \cite{tilt} that a flat domain wall
just above the depinning transition belongs to the universality class of Eq.~(\ref{ew}).
As mentioned above to come to this conclusion a number of assumptions have to be made.
The velocity of the domain wall must go to zero so that a KPZ-like quadratic term
of kinetic origin can be neglected, multiplicative noise must be assumed to be irrelevant
and the driving field must be replaced by an effective field which is the difference to the
field strength at the depinning transition. Since these assumptions are of heuristic
nature only it is of interest to study the underlying semi-microscopic model.
 
\section{Model}  
We study a model with non-conserved scalar order parameter (model A
in the classification of Hohenberg and Halperin \cite{HohHal}) with Langevin dynamics
at zero temperature,
\be
\gamma \frac{\partial \phi({\bf r},t)}{\partial t} 
 = - \frac{\partial {\cal{H}}}{\partial \phi({\bf r},t)} \, ,
\label{tdgl}
\ee
with a relaxation time proportional to $\gamma$. 
Here thermal noise is neglected since it is believed to be irrelevant \cite{zero}
at low temperatures.
The Ginzburg-Landau type Hamiltonian
$\cal{H}$ is given by
\be
{\cal{H}} =  \int d{\bf r}\, 
  \Bigl(-\frac{a}{2}\phi({\bf r},t)^2+\frac{b}{4}\phi({\bf r},t)^4
  +\frac{\tilde{J}}{2} \left(\nabla \phi({\bf r},t)\right)^2
  - (H+B({\bf r}))\,\phi({\bf r},t) \Bigr)
\label{gl}
\ee
where $\phi$ denotes a scalar order parameter, $H$ denotes a homogeneous
driving magnetic field and $B({\bf r})$ a quenched random field. It 
is assumed that the random fields have zero mean and are uncorrelated in space
and time. This
model and the following discretization are discussed in more detail
for the two dimensional or $d=1+1$ case in a previous paper \cite{mjt_pre}.

The discretization of Eqs.~(\ref{tdgl}) and (\ref{gl}) for the computational simulation 
in the present case $d=2+1$  is straight forward.
It results in a set of difference equations,
\be
S_{\bf l}(\tau+ \triangle \tau)  =  S_{\bf l}(\tau)+ \triangle \tau 
\biggl[ -u\left(S_{\bf l}^2(\tau)-1\right)S_{\bf l}(\tau)
 + \sum_{{\bf l}'}S_{{\bf l}'}(\tau)-6 S_{\bf l}(\tau)+h+b_{\bf l} \biggr] \, ,
\ee
which have to be iterated. 
The local magnetizations $S_{\bf l}$ may be termed
soft Ising spins at lattice point ${\bf l}=(x,y,z)$ with $-\infty<S_{\bf l}<\infty$.
The summation in Eq.~(4) is over the $6$ nearest neighbors $S_{\bf l}'$ 
of $S_{\bf l}$. 
The driving field $h$ and the quenched random fields $b_{\bf l}$ are measured in units 
of $\tilde{J}/{\delta^2}$, time $\tau$ is measured in units of 
$\gamma{\delta^2}/\tilde{J}$ and $u=a{\delta^2}/\tilde{J}=b{\delta^2}/\tilde{J}$
whereas $\delta$ denotes the lattice constant of the cubic lattice.
Eq.~(3) is iterated starting from a vertical flat initial interface where all spins 
on the left hand side of the interface located at $x=0$ are set to $S_{\bf l}=1$
and all spins at the right hand side are set to $S_{\bf l}=-1$. In $y$- and 
$z$-direction periodic boundary conditions are assumed. The random fields $b_{\bf l}$
are drawn with equal probability from an interval between $-p$ and $p$ with $p=0.6$,
$u$ is chosen as $u=0.9$ and the time constant as $\triangle \tau=0.1$.

\section{Results} 
The interface location
$\h({\bf r}_0,\tau)$ is defined as that point at which the magnetization 
$S_{\bf l}(\tau)$ as a function of $x$ for fixed ${\bf r}_0=(y,z)$ changes sign.
For not too large values of $p$ and $h$ the function $\h({\bf r}_0,\tau)$ is 
single valued since there are practically no overhangs or droplets.
Of interest are the height correlation function
\be
  C(r,\tau)=\overline{\langle [\h({\bf r}_0+{\bf r},\tau)-\h({\bf r}_0,\tau)]^2 \rangle} \, . 
\label{hcf}
\ee
and the width of the interface, which is often called the roughness,
\be
w(L,\tau)=\langle\overline{(\h({\bf r}_0,\tau) - \langle \h({\bf r}_0,\tau) \rangle)^2} \rangle^{1/2} 
\label{width}
\ee 
where the angular brackets denote an averaging over all interface sites and
the over-bar the average over different realizations of the quenched disorder.
Note that we have averaged here over ten configurations, an average over more 
configurations gave no better data.
These two functions are related to each other by 
\be
w(L,\tau)^2= \frac{1}{2L^{d-1}} \sum_{r=1}^{L} C(r,\tau) 
\ee
which is exact for periodic boundary conditions. 

Due to numerical limitations the largest possible system size was $L=128$.
Here we found that the depinning transition takes place at $h_C \simeq 0.0029$.
We have analyzed the morphology of the domain wall for a driving field slightly 
above this critical field $h=0.003$.
Fig.~(1a) shows the height correlation function as a function of distance $r$ for 
different times on logarithmic scales. 
For small $r$ a linear behavior is observed, i.e. $C(r,\tau)$ shows power law behavior,
while for large $r$ saturation sets in at $C$-values which depend on time. 
For this behavior which is well known
from kinetic roughening phenomena usually a dynamical scaling form \cite{Fam}
\be 
C(r,\tau) =  \xi(\tau)^{2\alpha} g\left(\frac{r}{\xi(\tau)}\right) 
\label{dsc}
\ee
works well. Here $\xi(\tau) \propto \tau^{1/z}$ denotes a time dependent correlation 
length and $z=\alpha/{\beta}$ a dynamic exponent. The scaling function $g(x)$ 
has the properties $g(x) \approx const$ for $x\gg 1$ and 
$g(x) \propto x^{2\alpha}$ for $x \ll 1$. 
The above scaling form contains two important limiting cases. 
The first one, 
\be
C(r\gg\xi(\tau),\tau) \propto \tau^{2\beta} \,,
\label{sbeta}
\ee
describes the growing of the spatially uncorrelated interface fluctuations with time
whereas the second one describes the growing of the spatial correlations:
\be
C(r\ll\xi(\tau),\tau) \propto r^{2\alpha} \,.
\label{salpha}
\ee
Analyzing the height correlation function in the above limiting cases we 
get $\alpha=0.68 \pm 0.01$ for the roughness exponent and for the small 
time exponent $\beta=0.36 \pm 0.02$ (see inset of Fig.~(1a)).
Note that we have performed simulations for different strength $p$ of the 
random fields and found no dependence of the exponents on $p$.
Due to the occurrence of an intrinsic width \cite{mjt_pre} 
of the interface, which manifests it selves in a turning point of the height 
correlation function for small $r$, the scaling form Eq.~(\ref{dsc}) holds only 
for $r \geq 3$. 
The average step height of the wall, characterized by $C(1,\tau)$, shows
a small time dependence, see Fig.~(1a). 
If we fit $C(1,\tau)$ to a power law, $C(1,\tau) \propto \tau^{2\kappa}$, as
we have done in $d=1+1$ \cite{mjt_pre} we find a very small value for this exponent,
$\kappa \approx 0.014$, on a time interval which is smaller than the
interval at which the scaling behavior Eq.~(\ref{sbeta}) is valid. 
Because of the smallness of this exponent and due to the small time interval where
an algebraic behavior can be observed we think that it makes not much sense to extend
the scaling law Eq.~(\ref{dsc}) as had to be done in $d=1+1$ where this
exponent is much larger \cite{mjt_pre}. Fig.~(1b) shows a satisfactory scaling plot.

Since in the limit $u \rightarrow \infty$ the present soft-spin model 
goes over into the random-field Ising model (RFIM) \cite{Binder} 
it is tempting to compare with this model.
The theoretical prediction of Grinstein and Ma \cite{GM}
for the RFIM is $\alpha=(5-d)/3$ while Ji and Robbins \cite{JR} obtained
from numerical studies of the RFIM in $d=3$ the value $\alpha=0.67 \pm0.03$.
Amaral et.al. \cite{tilt} found that for the RFIM the prefactor of the KPZ-nonlinearity 
is zero or goes to zero at the depinning transition so that the  RFIM
should belong 
to the EW-universality class. The value of the roughness exponent we obtain
is close to the above cited values. Note that Leschhorn \cite{Lesch} found for a
solid-on-solid (SOS)-type modification Eq.~(\ref{ew}) $\alpha=0.75 \pm
0.02$ significantly  
larger than our value. 

With the values for the roughness exponent $\alpha$ and the small time exponent $\beta$
we obtain for the dynamic exponent $z=\alpha/\beta=1.89 \pm 0.13$.
We consider this value as a
confirmation of our dynamic scaling analysis for Eq.~(\ref{ew}) 
presented earlier \cite{ROMA} where we found $z=2$ for all $d$ and $\alpha=(5-d)/3$.
Note that $z\approx 2$ was found previously also in 
$d=1+1$ for the model discussed here \cite{mjt_pre} and again in
$d=1+1$ from a direct integration of the equation of motion 
Eq.~(\ref{ew}) \cite{ROMA}.\\
The result for the dynamic exponent and therefore for the small
time exponent $\beta=\alpha/z$ disagrees with the numerical results of
Leschhorn for his SOS-model 
who found $\beta=0.475 \pm 0.015$ and $z=1.56 \pm 0.06$ in $d=3$. The
reason for the discrepancy of the numerical values of Leschhorn and
others is not clear to us especially since
Amaral et.al. \cite{tilt} could show that Leschhorns  
SOS model can also be described by Eq.~(\ref{ew}). 
A renormalization group
study of Eq.~(\ref{ew}) has been done by Nattermann et al. \cite{Natter} who found 
$z=2-2\epsilon/9$ in $d=5-\epsilon$. If one extrapolates this down to
$d=3$ which, however, might be well outside the range of the linear
${\epsilon}$-expansion one obtains rather good agreement with the value
of Leschhorn but not with our result.  Still further work
has to be done to clarify this situation.\\
The situation described so far is only valid for systems near the depinning 
transition. Far away from it, i.e. for driving fields $h \gg h_C$ the domain wall moves 
with a large velocity and the quenched random fields act as an effective white noise. 
Additionally, a nonlinear KPZ term is important \cite{KrugSpohn}.
Simulations in this regime are extremely difficult if not impossible
since one has to go to very large system sizes. For small systems
EW-behavior is observed and the crossover to KPZ-behavior occurs only
for very large system sizes in $d=1+1$ and for system sizes practically
outside the range for simulations in  $d=2+1$ \cite{Krug,NatterTang,ROMA}.

\pagebreak
\begin{figure}
 \epsfxsize=14.4cm
 \epsfysize=10.8cm
 \epsffile{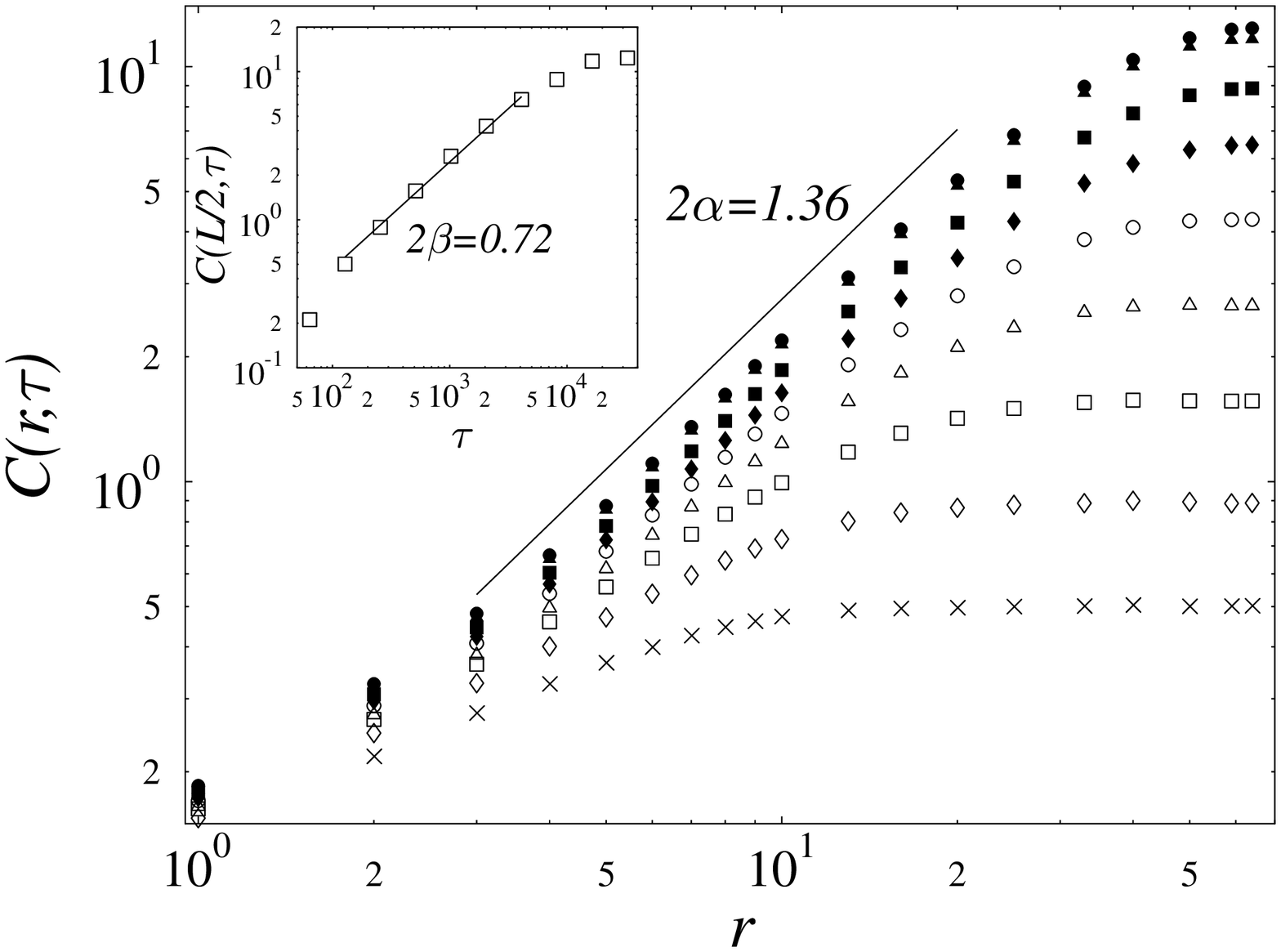} 

 \epsfxsize=14.4cm
 \epsfysize=10.8cm
 \epsffile{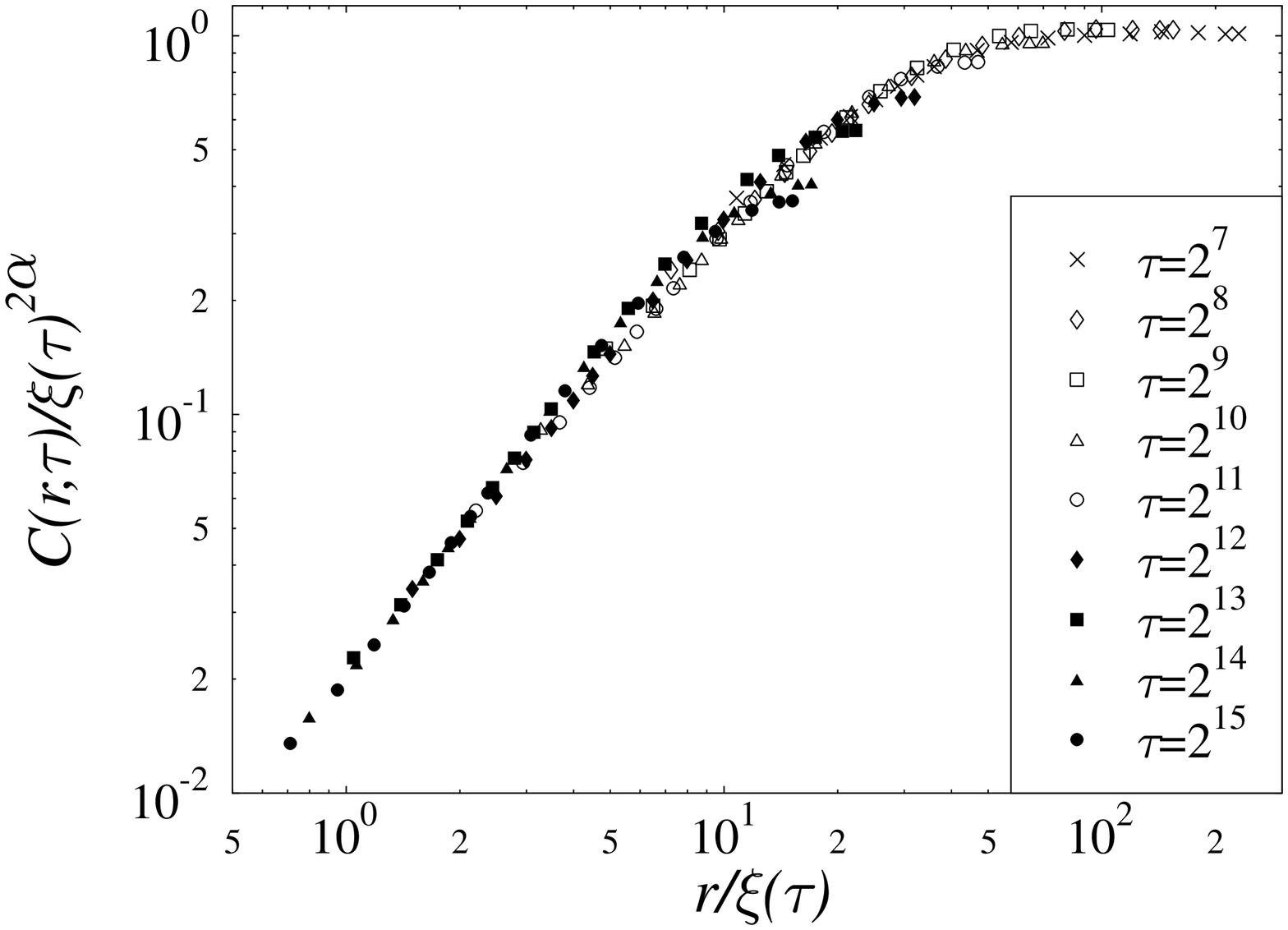}

\caption{ a) The height correlation function $C(r,\tau)$ for $h=0.003$ at various times as indicated
  in Fig.~(1b). The line represents the fit to
  $C(r,\tau \rightarrow \infty) \propto r^{2\alpha}$ (Eq.~(\ref{salpha})). 
  The inset shows the behavior of $C(L/2,\tau)$. 
  The line represents the fit $C(L/2,\tau) \propto \tau^{2\beta}$ (Eq.~(\ref{sbeta})).
 b) Scaling plot of the height correlation function according to Eq.~(\ref{dsc}) for $r \geq 3$ 
  with the same driving field as in (a).
}
\end{figure}

\end{document}